# CoCoFormer: A controllable feature-rich polyphonic music generation method

Jiuyang Zhou[0009-0008-3290-4801], Tengfei Niu[0009-0007-0322-0497], Hong Zhu[✉],

and Xingping Wang

Xi'an University of Technology, Xi'an. 710048, China
{2210320121, 2220321230}@stu.xaut.edu.cn, {zhuhong,
wangxingping}@xaut.edu.cn

**Abstract.** This paper explores the modeling method of polyphonic music sequence. Due to the great potential of Transformer models in music generation, controllable music generation is receiving more attention. In the task of polyphonic music, current controllable generation research focuses on controlling the generation of chords but lacks precise adjustment for the controllable generation of choral music textures. This paper proposes a Condition Choir Transformer (CoCoFormer) which controls the model's output by controlling the input of the chord and rhythm at a fine-grained level. This paper's self-supervised method improves the loss function and performs joint training through conditional control input and unconditional input training. This paper adds an adversarial training method to alleviate the lack of diversity in generated samples caused by teacher-forcing training. CoCoFormer enhances model performance with explicit and implicit inputs to chords and rhythms. In this paper, the experiments show that CoCoFormer has reached much better performance when compared to existing approaches. Based on specifying the polyphonic music texture, the same melody can also be generated in various ways.

## 1    INTRODUCTION

Music can be seen as an ordered combination of notes. According to music theory, notes are combined into parts, and multi-part notes constitute a rich musical texture. The temporal and spatial characteristics contained in a piece of music are defined as the textures of the music. According to different textures, music is divided into monophonic music, homophonic and polyphonic. Monophonic music contains only one melody line, which is the form of ancient music and most folk songs. Homophonic has a prominent melody line; other parts are subordinate to the main melody, such as harmony and accompaniment. Polyphonic music consists of several independent melodies. The lines are combined to reflect the characteristics of polyphonic music through different harmonies and polyphonic techniques. From the perspective of texture, the texture of monophonic music contains melody, the texture of homophonic music has horizontal melody and vertical harmony, and the texture of polyphonic music includes two forms of polyphony and harmony, as shown in Fig. 1.

Computer modeling of polyphonic music has been studied for decades, starting in the 1960s[1]. In recent years, much of the progress in music generative models has been due to the continued development and application of neural networks. Models in natural language processing have been widely used in music generation. Sequence modeling has always been the standard choice for music modeling, from early hidden Markov models to RNN[2], LSTM[3], BiLSTM[4], and other RNN methods; in addition to sequence modeling, a piece of music can be converted into a piano roll, and images can also be input. CNN trained as a generative adversarial network for music generation[5]. Some recent work revolves around Transformer[6]. Music Transformer[7] proposes a new relative attention mechanism and reduces the complexity of the model to linear as the sequence increases. MuseNet[8] uses a Transformer model based on GPT-2[9], which has been used in an amount of Music training can generate various styles of music such as classical and jazz.

The texture of polyphonic music can be seen as a combination of horizontal melodic features and vertical chord features. Chord features can be used as additional input to improve the generation quality of music models. The input in [10] has pitch and chord information, uses the expert system of the LSTM model[11] to generate jazz, MidiNet[12] uses CNN combined with GAN to generate music, and adds chord sequences for auxiliary prediction. But none of these methods can predict chords as input for each step, but treat them as fixed input.

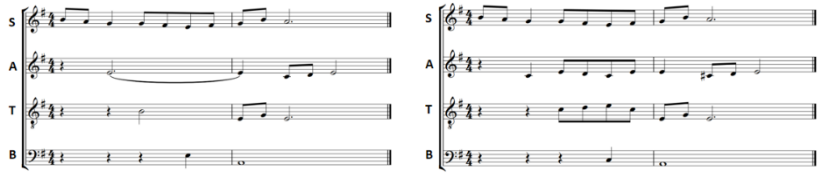

**Fig. 1.** The embodiment of texture in polyphonic music: the horizontal melody structure and the vertical harmony structure constitute the texture, which is embodied in specific rhythm and pitch. The left and right pictures use the same main melody, but the texture characteristics are entirely different, and people's hearing is also different.

The research on the harmony of polyphonic music has always taken the choral hymn data set of Johann Sebastian Bach as the research object[13][14], and the research goal is to generate polyphonic music that is like Bach as possible. For example, BachBot[15] uses the LSTM structure to generate choral music, DeepBach[16] uses DeepRNN for feature extraction and has been used in Monteverdi's five-part madrigal and Palestrina's mass. TonicNet[17] uses chord and pitch information and the number of repetitions of notes as data input, using GRU[18] for feature extraction. DeepChoir used the stacking structure of BiLSTM for sequence modeling and used the harmony consistency evaluation method of chords and melody to evaluate the results to generate chorus musicality.

Polyphonic music has two expression techniques: polyphony and harmony, embodying harmony and rhythm. The chords and rhythm of the melody reflect the textural characteristics of polyphonic music. This paper starts from the perspective of controllable music generation of polyphonic music textures, uses a self-attention mechanism to build a polyphonic music generation model CoCoFormer (Condition-Choir-Transformer), and selects Bach's choral hymns as a training sample set. This paper proposes a way to combine additional information explicitly and implicitly into the model and proposes a self-supervised learning method to construct a loss function for better training. Experiments have proven that CoCoFormer has achieved the best performance of known polyphonic music generation models.

The contributions of this paper are:

1. This paper propose a polyphonic music generation model CoCoFormer based on texture features, which can generate controllable choral music and proves the effectiveness of each module.

2. Introduce a self-supervised learning method, using multiple loss functions for joint training.

3. Through experiments, this paper studies how CoCoFormer controls music generation by controlling texture characteristics and generates music samples with precise control.

## 2    RELATED WORK

Recursive neural networks have potent capabilities for modeling sequence data. RNN-based models[19] have been widely used in music generation. The earliest example is to use RNN[20][21] to simulate single-note music. LSTM and other methods alleviate long-term dependence based on RNN models, replacing RNN as a more effective method.

In recent models, DeepBach adds extensions and beat marks while inputting pitch information. Through the forward DeepRNN[22] and reverse DeepRNN, the two modules perform bidirectional sequential extraction of features and output the probability of pitch. They are using pseudo-Gibbs sampling to simulate the process of artificial composition. The input of the TonicNet model includes word embeddings of chord and pitch information and word embeddings of input note repetition times, which are spliced and then input into the encoder. However, in generating long sequences, global and local information cannot be combined, and it is hard to learn the relationship between long-distance notes. The DeepChoir[23] model is a multi-head autoregressive encoding and decoding structure model. It uses the melody, chord information, rhythm, extension marks, and the notes of the soprano part as input and is encoded through a three-layer Bi-LSTM[24]. The output layer uses gamma sampling output.

In general, RNN-based models make the generated music more musical, and the additional information also improves the performance in generating polyphonic music. However, as the generated sequence continues to grow in application, it is difficult for the model to consider both global and local information to generate accurate notes.

Experiments show that the self-attention mechanism is significantly better than RNN-based models in establishing long-range dependencies[25], and its ability to extract semantic features is also higher than RNN-based models, indicating that Transformer is more suitable for music generation than RNN structures. For example, MusicTransformer adds relative position attention based on the structure of the Transformer and optimizes the time complexity of establishing long sequences. Lakh-NES[26] uses the structure of TransformerXL[27], is trained on a large four-part music library, and is evaluated on the NES dataset[28]. MuseNet is based on the GPT2[29] model and trained on many music data sets to predict the next token through unsupervised learning. In the direction of polyphonic music generation, Transformer models are rarely used in polyphonic music generation, and they also need more controllability of polyphonic music textures.

This paper proposes the CoCoFormer model. Because of the controllable problem of the current model in generating music, a polyphonic music generation method based on controllable chords and rhythm is proposed. Because of the auxiliary additional information, this paper designed a new structure to input a combination of chords and rhythms with a specific structure to extract the feature vectors of chords and rhythms and fuse them with the attention matrix to combine the vectors of polyphonic music explicitly and implicitly. This method enhances the controllability of the model texture.

## 3     PROPOSED MODEL COCOFORMER

Give a prompt message $x_{con} = \{c_1, \dots, c_n\}$ and note sequence $x = \{x_1, x_2, \dots, x_T\}$, joint probability $P(x|x_{con}) = \prod_t P(x_t|x_{<t}, x_{con})$. The task of this paper is to generate a model through an autoregression model, learn the factorization formula of $P(x|x_{con})$ distribution through an autoregressive generative model, encode the polyphonic music into a fixed-size feature matrix, and decode it through softmax to get the output of the next token.

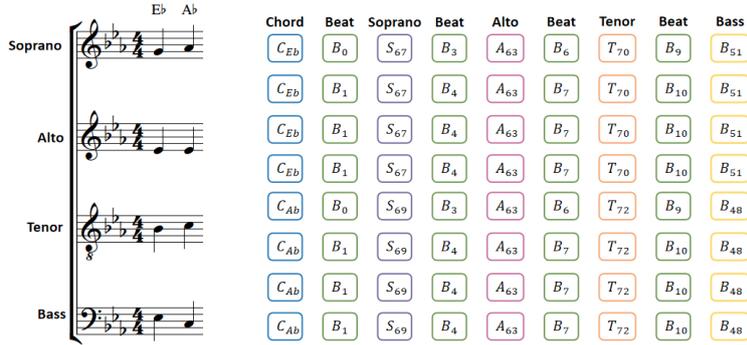

**Fig. 2.** Data representation of notes is arranged in sixteenth note resolution.

This paper will introduce CoCoFormer through data representation, model structure, and loss function construction.

### 3.1 Data Representation

In symbolic music generation, the pitch, chord, rhythm, and other information of each note are represented by specific events. This paper first defines the representation of events. Referring to the midi coding and defining 128 pitch events. For chord events, 49 types of events given in the data set are used, including 12 types of major chords, minor chords, augmented chords, diminished chords, and a category of other chords. Rhythm events give three states of notes under different voices: note start, rest, and note hold, a total of 12 types of events.

Given a symbol sequence of four parts of polyphonic music, and define the relevant markers: note marks: note $= \{N_s^0, N_a^0, N_t^0, N_b^0, \ldots, N_s^n, N_a^n, N_t^n, N_b^n\}$ , chord marks: chord $= \{C^0, C^1, C^2, \ldots, C^n\}$ , rhythm marks: beat $= \{B_s^0, B_a^0, B_t^0, B_b^0, \ldots, B_s^n, B_a^n, B_t^n, B_b^n\}$, as shown in Fig. 2.

The data input is expressed as：

$$\text{Condition input}_{\text{chord}} = \{C^0, C^1, C^2, \ldots, C^n\} \tag{1}$$

$$\text{Condition input}_{\text{beat}} = \{B_s^0, B_a^0, B_t^0, B_b^0, \ldots, B_s^n, B_a^n, B_t^n, B_b^n\} \tag{2}$$

$$\text{Input} = \{C^0, B_s^0, N_s^0, B_a^0, N_a^0, B_t^0, N_t^0, B_b^0, N_b^0, \ldots, C^n, B_s^n, N_s^n, B_a^n, N_a^n, B_t^n, N_t^n, B_b^n, N_b^n\} \tag{3}$$

The superscript of the mark is time, the sequence length is n, and the subscripts of rhythm mark and pitch mark represent different parts, s, a, t and b, respectively representing female high, female low, male high, and male low. Department, superscript indicates time.

### 3.2 CoCoFormer

This paper starts from the perspective of controllable texture and proposes the Co-CoFormer model through explicit and implicit input of chords and rhythms. As shown in Fig. 3, in the first stage, two single-layer standard Transformer encoders containing attention layers and feed-forward layers are designed to learn implicit expressions of rhythm and chords and concatenate them into the Transformer in stage two. The features are concatenated with the attention layer of the first Transformer layer. After the subsequent Transformer layers in the second stage, an attention layer that adds relative positions is used to learn the relationship between notes. The features are decoded to generate polyphonic music.

From equations (1) and (2), through token embedding and position embedding, the input chord and rhythm sequences are obtained $x_{\text{chord}} \in R^{l_c \times d_p}$, $x_{\text{beat}} \in R^{l_b \times d_p}$, where $d_{(\cdot)}$, $l_c$, $l_b$ respectively represent the feature dimension, the length of the chord sequence, and the length of the rhythm sequence. $H_{\text{chord}} \in R^{l_c \times d_p}$, $h_{\text{beat}} \in R^{l_b \times d_p}$ are obtained through two Transformers $\text{Model}_{\text{chord}}$, $\text{Model}_{\text{beat}}$ with the same structure but different sizes, such as (4)(5):

$$H_{\text{chord}} = \text{Model}_{\text{chord}}(x_{\text{chord}}) \tag{4}$$

$$H_{\text{beat}} = \text{Model}_{\text{beat}}(x_{\text{beat}}) \tag{5}$$

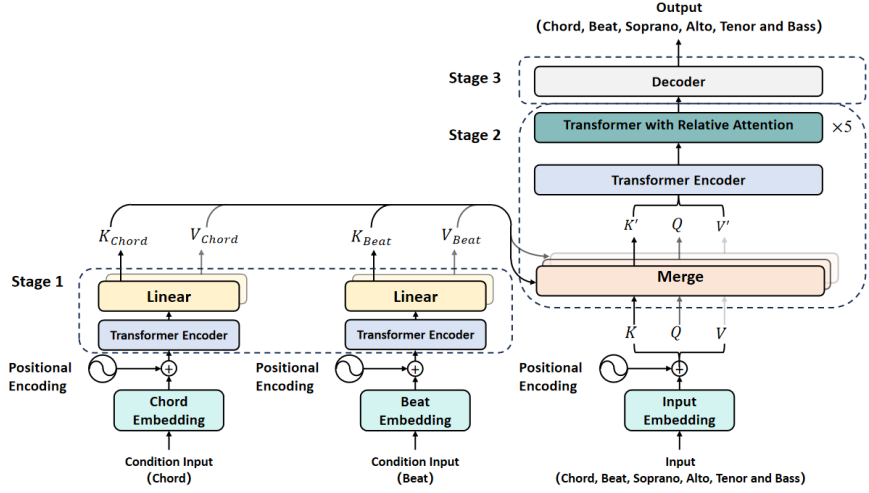

**Fig. 3.** The structure of CoCoFormer is divided into three parts: conditional input in Stage 1, feature extraction in Stage 2, and decoding in Stage 3

After $h_{chord}$, $h_{beat}$ passes through the linear layer, the matrix size is calculated unchanged with the corresponding $K_{chord}$, $V_{chord}$, $K_{beat}$, and $V_{beat}$. Then, concatenate features with input, such as (6), and calculate the attention and feed-forward layers as (7):

$$K' = [K_{chord}, K_{beat}, K], \ V' = [V_{chord}, V_{beat}, V] \tag{6}$$

$$A = \text{Softmax}(QK'^T)V', h_{out1} = \text{FFN}(A) \tag{7}$$

The backbone Transformer in the second stage is used to process polyphonic music features that add chords and rhythms. It is composed of the Decoder with attention adding relative position relationships, such as equation (8), where $R_I$ represents the relative position relationship.

$$\text{Head}_i = \text{RA}(Q_I, K_I, V_I, R_i) = \text{Softmax}(Q_i K_i^T + Q_i R_i)V_I \tag{8}$$

$$\text{Multi} - \text{RA}(Q, K, V) = \text{Concat}(\text{head}_1, \dots, \text{head}_n)W^o \tag{9}$$

In the third stage, the Decoder is constructed from three linear layers with normalization and dropout. The features are processed by the linear layer, and the probability of the next token is generated through softmax output:

$$P_t = \text{Softmax}(h_{out3\,(<t)}) \tag{10}$$

### 3.3    Loss Function

We train CoCoFormer through a self-supervised method, as shown in Fig. 4. The model input is divided into conditional input and primary input. Naturally, this paper uses different inputs to construct loss functions.

First, use the self-reconstruction method for training. Establish the joint probability $P(x_t) = \prod_t P(x_t | x_{<t}, c = x_{c<t}, x_{b<t})$ of input sequence $x = \{x_1, x_2, ..., x_T\}$. X is the conditional input of the model. Under the premise of given conditional input $x_{con}$, learn the distribution of the sequence, and the loss function constructed is as (11):

$$L_{self} = -\sum_{i=1}^{l} p_i \log_2\big(q_i(x_i | c = x_c, x_b)\big) \tag{11}$$

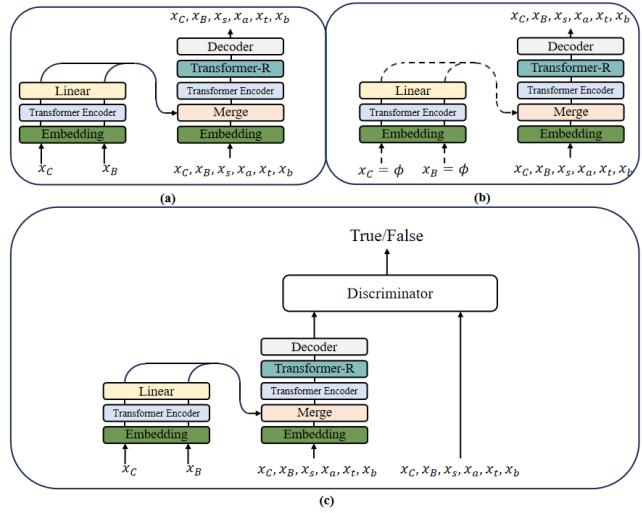

**Fig. 4.** CoCoFormer training method. (a) represents the training process of constructing a self-loss function, using conditional input and primary input for generation training. (b) represents generation training when the conditional input is blank. (c) constructs a discriminator to judge the music generated by the model.

To enable CoCoFormer to generate music smoothly without relying on conditional input, this paper uses a loss function similar to (10), but only uses input, set $input_{chord} = \emptyset$ and $input_{beat} = \emptyset$, as (12):

$$L_{null} = -\sum_{i=1}^{l} p_i \, log_2\big(q_i(x_i)\big) \tag{12}$$

The transformer is trained in a teacher-forced manner, which speeds up the convergence of the model. However, it is subject to ground-truth constraints. It is hoped that the results will correspond to the samples one-to-one, but on the other hand, it also reduces the possibility of generating diversity[30].

Adversarial training objectives have been shown to help generate sample output[31]. Based on this, this paper uses an adversarial training[32] loss function to minimize the loss to facilitate matching the output to the training samples:

$$L_{adv} = -E_x[\log f_{disc}(x)] + E_y[\log(1 - f_{disc}(y))] \tag{13}$$

$f_{disc}$ is a discriminator network consisting of a single layer of Transformer. They are used to determine whether the sample is a sample generated by the model. During the backpropagation process, the parameters of CoCoFormer $f_{disc}$ are updated, and the training goal is to minimize the loss function. Parameterize $f_{disc}$ with $\varphi$ , then the optimization objective is:

$$\Phi^* = \mathrm{argmin}_\varphi L_{adv} \tag{14}$$

The entire training process is trained through the linear combination of equations (11) (12) (13). CoCoFormer uses the Adam optimizer to optimize the following total loss function:

$$L = \mathrm{argmin}_\theta\big(\lambda_{self}L_{self} + \lambda_{null}L_{null} + \lambda_{adv}L_{adv}\big) \tag{15}$$

**Table 1.** COCOFORMER ABLATION EXPERIMENTAL RESULTS, USING VERIFICATION SET ACCURACY AS A METRICS

| Component | Choice | | | | | |
|---|---|---|---|---|---|---|
| Baseline | √ | √ | √ | √ | √ | √ |
| Chord cond | | √ | √ | √ | √ | √ |
| Relative attn | | | √ | √ | √ | √ |
| Rhythm cond | | | | √ | √ | √ |
| Loss1 | √ | √ | √ | √ | √ | √ |
| Loss2 | | | | | √ | √ |
| Loss3 | | | | | | √ |
| Accuracy | 0.8615 | 0.8726 | 0.913 | 0.932 | 0.9352 | 0.9404 |

## 4    EXPERIMENT

### 4.1    Dataset

This paper selects the JS Fake Chorales data set[33], selected from Bach's hymns with expanding similar music, and manually annotates chord information. The data set contains 500 four-part polyphonic music, with 49 events in the chord part, including 12 major chords, minor chords, augmented chords, diminished chords, and other chords. Note resolution is 16th note. To study the performance of the model on the original data set, this paper did not design any data enhancement method.

### 4.2    Experimental Setup

In the experiments, in the first stage, the Transformer used to process chords and rhythms which is a single layer, the input lengths are 256 and 1024 respectively, attention lengths are 8, the dimension of the middle layer of the feedforward layer is 256. Attention matrix in stage two is masked, and the Transformer layer in stage one is not masked. The input length of the Transformer in the second stage is 2048, the attention heads are 8, the dimension of the middle layer of feedforward is 1024, and the dropout is 0.1. The discriminator consists of a linear layer and a single layer of Transformer with 4 attention heads, the dimension of the middle layer of feedforward is 512, and the dropout is 0.5. The model training uses a GTX 1080Ti, and the training time is about 5 hours.

### 4.3    Ablation Test

This paper conducts ablation experiments on CoCoFormer and experiments on the structures added to the model and different data representation methods. This paper uses the highest accuracy of the validation set as a metric. As shown in Table 1, baseline represents the Transformer baseline, and Chord cond describes adding conditional chord information and the corresponding network. Relative attn represents using relative positional attention in the second stage. Rhythm cond means adding conditional rhythm information and the related processing network. The loss functions Loss1 and Loss2 use the cross-entropy function to process the generated samples of conditional input and unconditional input. Loss3 is the loss function of the adversarial network.

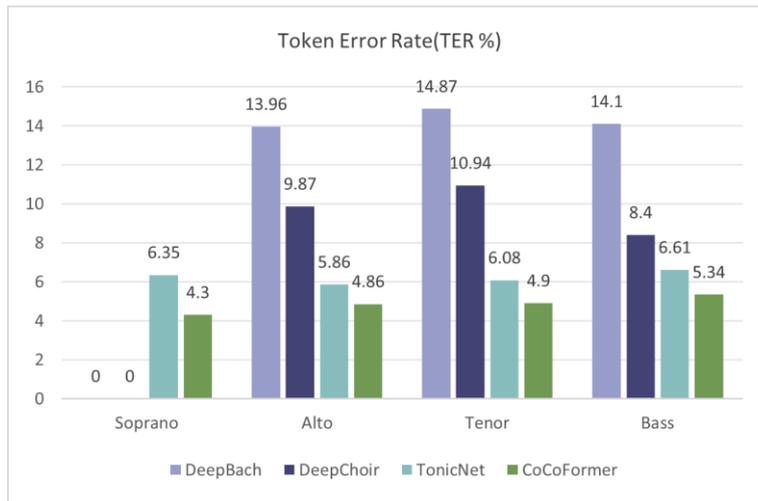

**Fig. 5.** CoCoFormer compares the token error rates of several current mainstream models.

Experiments have proven the effectiveness of improving the model: after adding the conditional chord model, the accuracy of the validation set increased by 1.11%, indicating that the model improved the quality of generated samples by adding extra information. Compared with the model that used relative positional attention in the second stage, the improved accuracy increased by 4.04%, indicating that chords combined with relative positions play a more obvious guiding role in note generation. The addition of extra information makes the sample generation closer to the data set samples, adding the rhythm information of the notes also conforms to this rule.

Improving the loss function is mainly aimed at generating samples more smoothly. Training under both unconditional and conditional inputs can enhance the stability of the model. Secondly, in order to alleviate the impact of teacher-forcing training on the diversity of sample generation, this paper set an adversarial loss function in training. Experiments have proven that using three loss functions for joint training can learn the potential expression of polyphonic music more effectively.

## 4.4     Comparative Test

This paper selects the main models currently used for polyphonic music generation and makes statistics on the token error rates(TER) of each part generation. As shown in Fig.5. Due to the structure of the model, the soprano parts of DeepBach and DeepChoir are not involved in the generation and are set to 0. CoCoFormer achieved the best performance of these current models in the comparison results.

DeepBach, DeepChoir, and TonicNet are all models with RNN-based. Experiments proved that Transformer performs better in data set evaluation after solving the long-term dependency problem. It is worth noting that training samples of TonicNet expanded the data set to 1968 polyphonic music by transposing all pieces as many samples as possible, while CoCoFormer did not use any data augmentation methods.

## 4.5     Qualitative Experiment

Generating polyphonic music through specified chords and rhythms can make the generated music more suitable for different application scenarios and according to user needs. In the previous models, BachBot and TonicNet did not explicitly use rhythm condition as input, which made the rhythm generated uncontrollable. DeepBach and DeepChoir placed the duration mark of the note in the pitch event, and the beats events are related to the time signatures and beats of the generated music. (The rhythm of music input in 4/4 beat is strong beat, weak beat, sub-strong beat and weak beat). Such a rhythm event setting makes it difficult for the input model to control the rhythm of a single note. The rhythm coding proposed in this article includes the rest, extension and starting process of notes, so the generated music can be controlled at the note level.

This paper conducted experiments using the same melody input and different rhythm and chord conditions to analyze the output. As shown in Fig. 6, this paper performed a visually analyzed of the piano roll axis on an input 8-bar music. The main melody of the input remains unchanged. Only the texture is changed. The first line changes the input chord information and keeps the rhythm information unchanged. The experiment has shown that output notes keep the rhythm unchanged, and the harmony changes. The first column aligns the rhythm generated by the other three parts with the rhythm of the input melody, keeping the chord information unchanged and changing the rhythm of the generated sample texture. Experiments have proven that CoCoFormer can generate diversity based on given prompt information so that the same input melody can generate polyphonic music with different textures.

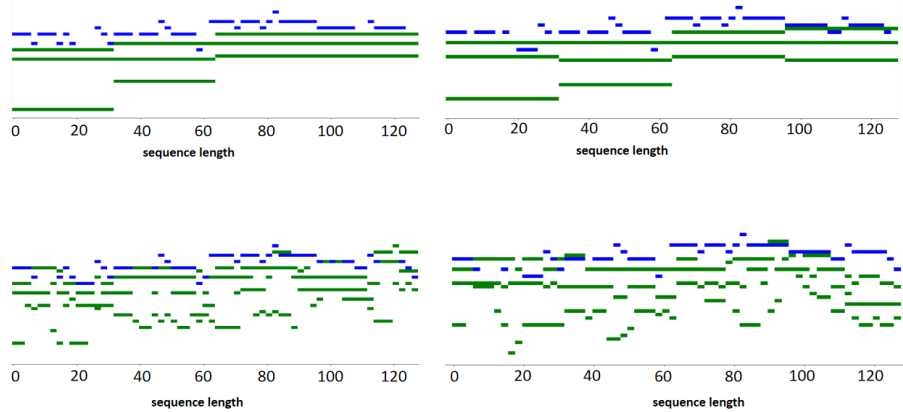

**Fig. 6.** Specify a melody and change the input chords and rhythm while keeping the melody. The upper left uses the melody's original chord and rhythm input, and the upper right changes the chord encoding while keeping the rhythm input unchanged. In the lower left, the input rhythm is consistent with the rhythm of the input melody, and the chords are unchanged. In the lower right, the chords and rhythm are also changed.

Fig. 7 conducts a comparative experiment on whether to use the second stage of conditional encoding, gives an example of the score generated by Jasmine's chorus music. This paper specifies four voices to maintain a consistent rhythm. It can be seen that after adding conditional encoding, the generated music maintains better rhythmic and harmony consistency. In [34] we provide more music samples generated by Co-CoFormer.

**Fig. 7.** The score of CoCoFormer under the specified conditional encoding with the melody of Jasmine and rhythm of are consistent. The left picture shows the generation using the second stage conditional coding. The right picture leaves the input conditional coding empty and only uses the backbone network for generation. Compared with the right picture, the left picture maintains a higher consistency in rhythm (the boxed part of the left picture), and the right picture has more wrong harmonies in the chord generation (the boxed part of the right picture)

## 5    CONCLUSIONS

Polyphonic music generation remains a challenging problem. This paper proposes Co-CoFormer, which can not only better model polyphonic music but also have more controllable texture generation. The model proposes a new attention mechanism so that chord and rhythm information can be explicitly and implicitly extracted by features. The loss function proposed in this paper is optimized through a combination of cross entropy and adversarial loss, and the network is trained to generate under given conditions and unconditionally, which improves the coherence of the generated melody. Experiments have proven that CoCoFormer can dynamically adjust the texture composition according to specified requirements and generate polyphonic music of different styles, showing good generalization performance.